\documentclass[a4paper,12pt]{article}

\usepackage{graphics, epsfig}

\begin{document}

\author{P A. Hogan\thanks{E-mail : peter.hogan@ucd.ie}
and S. O'Farrell\thanks{E-mail : shanefarrell\_ie@hotmail.com}\\
\small School of Physics,\\ \small University College Dublin,
Belfield, Dublin 4, Ireland}

\title{Modelling Background Radiation in Isotropic Cosmologies}
\date{}
\maketitle

\begin{abstract}
 Using explicit perturbations of isotropic cosmological models which describe 
simple gravitational waves, an isotropic tensor having the algebraic symmetries of the Bel--Robinson tensor is derived as a model of cosmic background 
gravitational radiation and this is used to provide an answer to the question: in what sense can 
an energy--momentum--stress tensor similar to that describing the cosmic microwave background radiation (neglecting anisotropies) 
be associated with an isotropic background of \emph{gravitational} radiation?

\end{abstract}
\thispagestyle{empty}
\newpage

\section{Introduction}\indent
It is well known (see, for example, \cite{MTW}) that, neglecting small anisotropies, the cosmic microwave background radiation (CMB) in the universe is well described by a homogeneous perfect fluid which shares 
its 4--velocity with that of the isotropic matter distribution and has an equation of state giving the isotropic pressure equal to one third of the proper energy density. This 
paper is concerned with providing an answer to the question: \emph{in what sense is an isotropic background of gravitational radiation described by a similar perfect fluid to that of the CMB?} To answer this question 
we make use of (a) simple explicit gravitational waves described as small perturbations of Friedmann--Lema\^itre--Robertson--Walker (FLRW) cosmological models and (b) the Bel--Robinson tensor \cite{PR}. In section 2 we 
introduce notation and describe a derivation of the perfect fluid energy--momentum--stress tensor for the CMB in a way which motivates our approach. In section 3 the corresponding calculation is performed on the Bel--Robinson tensor and in section 4 a derivation of a perfect fluid energy--momentum--stress tensor 
for a background of gravitational radiation is given.
\setcounter{equation}{0}
\section{Cosmic Background Electromagnetic Radiation}\indent
The metric tensor components $(g_{ab})$ corresponding to the FLRW cosmological models are given via the well--known Robertson--Walker line--element
\begin{equation}\label{2.1}
ds^2=g_{ab}dx^a\,dx^b=R^2(t)\,f^{-2}\{dx^2+dy^2+dz^2\}-dt^2\ ,\end{equation}
with $R(t)$ the scale factor and 
\begin{equation}\label{2.1a}
f=1+\frac{k}{4}(x^2+y^2+z^2)\ ,\end{equation}
where $k=0, \pm 1$ is the Gaussian curvature of the homogeneous hypersurfaces $t={\rm constant}$. The matter 
content of these universe models is a perfect fluid with the integral curves of the vector field $\partial/\partial t=u^a\partial/\partial x^a$ the 
world lines of the fluid particles. Here $x^a=(x, y, z, t)$ for $a=1, 2, 3, 4$ respectively. The coordinate $t$ is proper time and $x, y, z$ are 
each constant along a fluid particle world line. The vector field with components $u^a$ satisfies $g_{ab}u^au^b=u^au_a=-1$ and is 
the 4--velocity of a fluid particle. The matter world lines are time--like geodesics, thus
\begin{equation}\label{2.2}
\dot u^a\equiv u^a{}_{;b}u^b=0\ ,\end{equation}with the semi--colon indicating covariant differentiation with respect to the Riemannian connection
associated with the metric tensor $g_{ab}$. Throughout this paper we follow the definitions and sign conventions of \cite{E1} and in particular a dot on 
a tensor will indicate covariant differentiation in the direction of $u^a$. These geodesic world lines have vanishing shear
\begin{equation}\label{2.3}
\sigma _{ab}\equiv u_{(a;b)}+\dot u_{(a}u_{b)}-\frac{1}{3}\theta\,h_{ab}\ ,\end{equation}with the round brackets denoting symmetrisation, $h_{ab}=g_{ab}+u_a\,u_b$
is the projection tensor and the scalar $\theta$ denotes the expansion of the world lines given by
\begin{equation}\label{2.4}
\theta \equiv u^a{}_{;a}=3\,\frac{\dot R}{R}\ .\end{equation}The dot on $R(t)$ denotes differentiation with respect to $t$. In addition the vorticity tensor
\begin{equation}\label{2.5}
\omega _{ab}\equiv u_{[a;b]}+\dot u_{[a}u_{b]}\ ,\end{equation}vanishes in the case above. The square brackets here denote skew symmetrisation. The energy--momentum--stress tensor of the matter is isotropic and 
thus has in general the perfect fluid form
\begin{equation}\label{2.6}
T^{ab}=\mu\,u^au^b+p\,h^{ab}\ ,\end{equation}with proper density $\mu$ and isotropic pressure $p$ functions of $t$ only (and thus satisfying $h^b_a\mu _{,b}=
0=h^b_ap_{,b}$). Einstein's field equations 
\begin{equation}\label{2.7}
R_{ab}=T_{ab}-\frac{1}{2}Tg_{ab}\ ,\end{equation}where $R_{ab}$ is the Ricci tensor obtained from the components of the Riemann curvature tensor 
according to the formula $R_{ab}=R^c{}_{acb}$, $T=T^a{}_a$ and a factor involving the gravitational constant has been absorbed into $T_{ab}$ for convenience, 
yield in the above case 
\begin{eqnarray}
p&=&-\frac{\dot R^2}{R^2}-\frac{2\ddot R}{R}-\frac{k}{R^2}\ ,\label{2.8a}\\
\mu &=&\frac{3\dot R^2}{R^2}+\frac{3k}{R^2}\ .\label{2.8b}\end{eqnarray}We see from these that 
\begin{equation}\label{2.9}
\mu +p=\frac{2}{R^2}(\dot R^2-R\ddot R+k)\ ,\end{equation}and so it easily follows from (\ref{2.8a}) and (\ref{2.8b}) that
\begin{eqnarray}
\dot\mu &=&-(\mu +p)\theta\ ,\label{2.10a}\\
\dot\theta &=&-\frac{1}{3}\theta ^2-\frac{1}{2}(\mu +3p)\ .\label{2.10b}\end{eqnarray}The first of these is the equation of conservation of energy along 
a fluid world line and the second is Raychaudhuri's equation.

Electromagnetic test fields on the space--times described above have electric and magnetic parts defined respectively by
\begin{equation}\label{2.11}
E_a=F_{ab}u^b\qquad {\rm and}\qquad H_a={}^*F_{ab}u^b\ ,\end{equation}where $F_{ab}=-F_{ba}$ is a tensor field on the 
FLRW space--time having dual ${}^*F_{ab}=\frac{1}{2}\eta _{ab}{}^{rs}F_{rs}$ with $\eta _{abcd}=\epsilon_{abcd}\sqrt{-g}$,  $g={\rm det}(g_{ab})$ 
and $\epsilon_{abcd}$ the four dimensional Levi--Civita permutation symbol. Maxwell's source--free field equations on the FLRW space--times 
read \cite{E1}
\begin{equation}\label{2.12}
E^a{}_{;a}=0\ ,\qquad H^a{}_{;a}=0\ ,\end{equation}and
\begin{eqnarray}
\dot E^a+\frac{2}{3}\theta\,E^a&=&-\eta ^{abed}\,u_b\,H_{e;d}\ ,\label{2.13a}\\
\dot H^a+\frac{2}{3}\theta\,H^a&=&\eta ^{abed}\,u_b\,E_{e;d}\ .\label{2.13b}\end{eqnarray}To solve these equations we introduce a 4--potential $\sigma ^a$ satisfying \cite{EH} 
\begin{equation}\label{2.14}
\sigma ^a\,u_a=0\ ,\qquad \sigma ^a{}_{;a}=0\ ,\end{equation}from which the Maxwell tensor $F_{ab}$ is calculated using
\begin{equation}\label{2.15}
F_{ab}=\sigma _{b;a}-\sigma _{a;b}\ .\end{equation}Now (\ref{2.12}) and (\ref{2.13b}) are automatically satisfied while (\ref{2.13a}) yields the wave 
equation (see \cite{EH}):
\begin{equation}\label{2.16}
\sigma ^{a;d}{}_{;d}=\frac{1}{2}(\mu -p)\,\sigma ^a\ .\end{equation} 

Some simple sinusoidal, monochromatic electromagnetic waves on a $k=0$ FLRW space--time satisfying (\ref{2.14}) and (\ref{2.16}) and thus (\ref{2.12})--(\ref{2.13b})
are constructed as follows (see \cite{HOF} for families of test electromagnetic radiation fields on isotropic cosmologies):  Let
\begin{equation}\label{2.17}
\phi =\delta_{\alpha\beta}\,n^{\alpha}\,x^{\beta}-T\ ,\end{equation}where $x^{\alpha}=(x, y, z)$ with Greek indices taking values $1, 2, 3$. Here $\delta_{\alpha\beta}$ is the three dimensional Kronecker delta and we note from (\ref{2.1}) that $\delta_{\alpha\beta}=R^{-2}g_{\alpha\beta}$. Also $n^{\alpha}$ are constants and $T$ is a function of $t$ only 
satisfying
\begin{equation}\label{2.18}
{\bf n}\cdot{\bf n}=\delta_{\alpha\beta}\,n^{\alpha}\,n^{\beta}=1\qquad {\rm and} \qquad \frac{dT}{dt}=\frac{1}{R(t)}\ ,\end{equation}respectively. With the metric tensor given via (\ref{2.1}) with $k=0$ it follows that 
\begin{equation}\label{2.19}
g^{ab}\phi _{,a}\,\phi _{,b}=0\ ,\end{equation} and so the hypersurfaces $\phi (x^{\alpha}, t)={\rm constant}$ are null hypersurfaces in the FLRW space--time with $k=0$. They 
will be the histories of the wave fronts of the electromagnetic radiation. As candidate solutions we take a potential 1--form 
\begin{equation}\label{2.20}
\sigma _a\,dx^a=\lambda\,\,\delta_{\alpha\beta}(B^{\beta}\cos\phi+C^{\beta}\sin\phi)\,dx^{\alpha}\ .\end{equation}Here ${\bf B}=(B^{\alpha})$ and ${\bf C}=(C^{\alpha})$ are constant 3--vectors related to 
the 3--vector ${\bf n}=(n^{\alpha})$ by
\begin{equation}\label{2.21}
{\bf B}={\bf b}\times {\bf n}\qquad{\rm and}\qquad{\bf C}={\bf b}-({\bf b}\cdot{\bf n})\,{\bf n}\ ,\end{equation}for any 3--vector ${\bf b}$ such that ${\bf b}\cdot{\bf b}=1$ and for any 
real constant $\lambda$. We note that 
\begin{equation}\label{2.22}
{\bf n}\cdot{\bf B}={\bf n}\cdot{\bf C}={\bf B}\cdot{\bf C}=0\qquad{\rm and}\qquad{\bf B}\cdot{\bf B}={\bf C}\cdot{\bf C}=1-({\bf b}\cdot{\bf n})^2\ ,\end{equation}and we have the useful relation
\begin{equation}\label{2.23}
\delta_{\mu\alpha}\delta_{\nu\beta}\left\{n^{\alpha}\,n^{\beta}+({\bf B}\cdot{\bf B})^{-1}(B^{\alpha}\,B^{\beta}+C^{\alpha}\,C^{\beta})\right\}=\delta _{\mu\nu}\ .\end{equation}To verify that (\ref{2.20}) satisfies (\ref{2.14}) and (\ref{2.16}) we note 
that each of the two terms on the right hand side of (\ref{2.20}) has the general form
\begin{equation}\label{2.24}
\sigma _a\,dx^a=s_a\,F(\phi)\,dx^a\ ,\end{equation}for any function $F$. Then the equations in (\ref{2.14}) imply \cite{EH}
\begin{equation}\label{2.25}
s^a\,u_a=0\ ,\qquad s^a{}_{;a}=0\qquad{\rm and}\qquad s^a\,\phi _{,a}=0\ ,\end{equation}while (\ref{2.16}) provides us with the equations: 
\begin{eqnarray}
g^{ab}\phi _{,a}\,\phi _{,b}&=&0\ ,\label{2.26a}\\
s^{a;b}\,\phi _{,b}+\frac{1}{2}\phi _{,d}{}^{;d},s^a&=&0\ ,\label{2.26b}\\
s^{a;d}{}_{;d}&=&\frac{1}{2}(\mu -p)\,s^a\ .\label{2.26c}\end{eqnarray}Now each choice of $s_a$ leading to (\ref{2.20}) is of the form
\begin{equation}\label{2.27}
s_a=(\delta_{\alpha\beta}\,a^{\beta}, 0)\qquad{\rm with}\qquad {\bf a}\cdot{\bf n}=0\ ,\end{equation}where ${\bf a}$ is a constant 3--vector and $\phi$ is given by (\ref{2.17}). 
Thus with the metric tensor given via (\ref{2.1}) with $k=0$ it easily follows that (\ref{2.25}) and (\ref{2.26a}) are satisfied. To verify that (\ref{2.26b}) and (\ref{2.26c}) 
are satisfied requires
\begin{equation}\label{2.28}
\phi _{,d}{}^{;d}=\frac{2\,\dot R}{R}\ ,\end{equation}$p$ and $\mu$ given by (\ref{2.8a}) and (\ref{2.8b}) and the non--vanishing components of the Riemannian 
connection associated with the metric $g_{ab}$ with $k=0$, namely, $\Gamma ^{\alpha}_{\beta 4}=\dot R/R\,\delta _{\alpha\beta}$ and $\Gamma ^4_{\alpha\beta}
=R\,\dot R\,\delta _{\alpha\beta}$. We note that the propagation direction in the FLRW space--time with $k=0$ of the histories of the waves is $\phi _{,a}$ and we see from
(\ref{2.28}) that the null geodesic integral curves of this covariant vector field have non--vanishing expansion on account of the expansion of the universe and so the waves 
are not plane waves. The unit 3--vector 
${\bf n}$ gives the direction of the wave propagation at any point in the 3--space $t={\rm constant}$. Substituting (\ref{2.20}) into (\ref{2.15}) and then calculating (\ref{2.11}) 
we find that $E_a=(E_{\alpha}, 0)$ and $H_a=(H_{\alpha}, 0)$ with
\begin{eqnarray}
E_{\alpha}&=&\frac{\lambda}{R}\,\delta_{\alpha\beta}(-B^{\beta}\sin\phi +C^{\beta}\cos\phi )\ ,\label{2.29a}\\
H_{\alpha}&=&-\frac{\lambda}{R}\,\delta_{\alpha\beta}(B^{\beta}\cos\phi +C^{\beta}\sin\phi )\ ,\label{2.29b}\end{eqnarray}
and in addition we have
\begin{equation}\label{2.30}
F_{\alpha 4}=-F_{4\alpha}=E_{\alpha}\qquad{\rm and}\qquad F_{\alpha\beta}=-F_{\beta\alpha}=R\,\epsilon _{\alpha\beta\gamma}\,H_{\gamma}\ ,\end{equation}
with $\epsilon _{\alpha\beta\gamma}$ the three dimensional Levi--Civita permutation symbol. The electromagnetic energy tensor associated with these 
electromagnetic waves is calculated from
\begin{equation}\label{2.31}
M_{ab}=\frac{1}{2}(F_{ac}\,F_b{}^c+{}^*F_{ac}\,{}^*F_b{}^c)\ .\end{equation}Here $M_{ab}=M_{ba}$ and $M^a{}_a=0$. Using (\ref{2.29a})--(\ref{2.30}) and making use of the relation (\ref{2.23}) (in particular to simplify 
$M_{\mu\nu}$ below) we arrive at
\begin{eqnarray}
M_{\mu\nu}&=&\frac{\lambda ^2}{R^2}({\bf B}\cdot{\bf B})\,\delta_{\mu\alpha}\,\delta_{\nu\beta}\,\,n^{\alpha}n^{\beta}\ ,\label{2.32a}\\
M_{\mu 4}&=&-\frac{\lambda ^2}{R^3}({\bf B}\cdot{\bf B})\,\delta_{\mu\alpha}\,n^{\alpha}\ ,\label{2.32b}\\
M_{44}&=&\frac{\lambda ^2}{R^4}({\bf B}\cdot{\bf B})\ .\label{2.32c}\end{eqnarray}These components can be summarized economically as
\begin{equation}\label{2.33}
M_{ab}=\frac{\lambda ^2}{R^2}({\bf B}\cdot{\bf B})\phi _{,a}\,\phi _{,b}\ ,\end{equation}with ${\bf B}\cdot{\bf B}$ given in terms of ${\bf b}$ 
and ${\bf n}$ by (\ref{2.22}). We now have explicitly how $M_{ab}$ depends upon the directions of ${\bf b}$ and ${\bf n}$. To obtain a model of the 
background electromagnetic radiation in the $k=0$ FLRW universe we sum the electromagnetic energy--momentum tensors for all 
such wave systems labelled by the directions of ${\bf b}$ and ${\bf n}$. Summing (i.e. integrating) $M_{ab}$ over the directions 
of ${\bf b}$ we find that
\begin{eqnarray}
M_{\mu\nu}&=&\frac{8\pi\lambda ^2}{3R^2}\delta_{\mu\alpha}\,\delta_{\nu\beta}\,\,n^{\alpha}n^{\beta}\ ,\label{2.34a}\\
M_{\mu 4}&=&-\frac{8\pi\lambda ^2}{3R^3}\,\delta_{\mu\alpha}\,n^{\alpha}\ ,\label{2.34b}\\
M_{44}&=&\frac{8\pi\lambda ^2}{3R^4}\ .\label{2.34c}\end{eqnarray}Now summing (i.e. integrating) over the directions of ${\bf n}$ we 
obtain the energy--momentum--stress tensor of the electromagnetic background radiation ${\cal M}_{ab}$ with components
\begin{eqnarray}
{\cal M}_{\mu\nu}&=&\frac{c_0^2}{R^2}\,\delta _{\mu\nu}\ ,\label{2.35a}\\
{\cal M}_{\mu 4}&=&0\ ,\label{2.35b}\\
{\cal M}_{44}&=&\frac{3c_0^2}{R^4}\ ,\label{2.35c}\end{eqnarray}where $c_0^2=32\pi ^2\lambda ^2/9$. This can be written in the form
\begin{equation}\label{2.36}
{\cal M}_{ab}=\mu_eu_a\,u_b+p_eh_{ab}\ ,\end{equation}
with
\begin{equation}\label{2.37}
p_e=\frac{1}{3}\mu _e=\frac{c_0^2}{R^4}\ ,\end{equation}and the subscript is used to reflect the electromagnetic origin of these quantities. Hence the model background electromagnetic radiation field that 
we have constructed in this way is a perfect fluid with fluid 4--velocity $u^a$ and equation of state (\ref{2.37}). We note that ${\cal M}^{ab}{}_{;b}=0$.

We now generalize the construction above to include the FLRW universes with $k=\pm 1$. The end result is expected to be again the 
isotropic energy--momentum--stress tensor (\ref{2.36}). Hence we can simplify the derivation by calculating this tensor along any one of 
the integral curves of the vector field $u^a$ and then extending the result to all curves on the basis of isotropy. For simplicity we work 
on the curve $r=0$ with $r^2=x^2+y^2+z^2$. We will establish that $s^a$ given by (\ref{2.27}), with $\phi$ given by (\ref{2.17}), in the neighbourhood 
of $r=0$ yields a solution of the equations (\ref{2.14}) and (\ref{2.16}) \emph{along} $r=0$.  This is quite a remarkable fact which we will see later 
is not replicated in the gravitational radiation case. With it established we then have that (\ref{2.36}) will hold along $r=0$ and hence 
by isotropy will hold everywhere in the FLRW space--time with $k=\pm 1$. We note that $\phi$ in (\ref{2.17}) satisfies (\ref{2.19}) 
\emph{along} $r=0$ with $g_{ab}$ given via (\ref{2.1}) and (\ref{2.1a}) with $k=\pm 1$. With $f$ given by (\ref{2.1a}) we have 
\emph{along} $r=0$:
\begin{equation}\label{2.38}
f=1\ ,\qquad \frac{\partial f}{\partial x^{\alpha}}=0\qquad{\rm and}\qquad\frac{\partial ^2f}{\partial x^\alpha\partial x^{\beta}}=
\frac{k}{2}\,\delta _{\alpha\beta}\ .\end{equation}We shall require the components of the Riemannian 
connection associated with the metric tensor $g_{ab}$ given via  (\ref{2.1}) and (\ref{2.1a}) with $k=\pm 1$. These are listed in the appendix. 
Since now $s^a=(f^2a^{\alpha}/R^2, 0)$ and $s_a=(\delta_{\alpha\beta}a^{\beta}, 0)$ we find that
\begin{equation}\label{2.39}
s^a_{;a}=\frac{2f}{R^2}\frac{\partial f}{\partial x^{\alpha}}\,s_{\alpha}+\frac{f^2}{R^2}\Gamma ^{\beta}_{\alpha\beta}\,s_{\alpha}\ ,\end{equation}
From this it follows that \emph{along} $r=0$ we have $s^a{}_{;a}=0$ and $(s^a{}_{;a})_{,b}u^b=0$ but $(s^a{}_{;a})_{,\lambda}\neq 0$. Using (\ref{2.38}) 
and the formulas in the appendix we find that along $r=0$ 
\begin{equation}\label{2.40}
(s^a{}_{;a})_{,\lambda}=\frac{2}{R^2}\frac{\partial ^2f}{\partial x^{\alpha}\partial x^{\lambda}}\,s_{\alpha}+\frac{1}{R^2}\Gamma ^{\beta}_{\alpha\beta ,\lambda}s_{\alpha}
=\frac{k}{R^2}s_{\lambda}-\frac{3k}{2R^2}s_{\lambda}=-\frac{k}{2R^2}s_{\lambda}\ .\end{equation}Now with $(s^a{}_{;a})_{,\lambda}\neq 0$ 
along $r=0$ the equations (\ref{2.25}), (\ref{2.26a}) and (\ref{2.26b}) hold (and are satisfied) along $r=0$. However the wave equation (2.32) is 
modified to read
\begin{equation}\label{2.41}
s_{a;d}{}^{;d}=\frac{1}{2}(\mu -p)\,s_a+h^b_a\,(s^d{}_{;d})_{,b}\ ,\end{equation}along $r=0$. This is trivially satisfied when $a=4$ and thus we must have 
\begin{equation}\label{2.42}
s_{\lambda ;d}{}^{;d}=\frac{1}{2}(\mu -p)\,s_{\lambda}+(s^d{}_{;d})_{,\lambda}\ ,\end{equation}along $r=0$. Direct calculation starting with $s_{\lambda}=a^{\lambda}$ and using 
the connection components given in the appendix reveals that along $r=0$
\begin{equation}\label{2.43}
s_{\lambda ;d}{}^{;d}=\left (\frac{\ddot R}{R}+\frac{2\dot R^2}{R^2}+\frac{3k}{2R^2}\right )\,s_{\lambda}=\frac{1}{2}(\mu -p)s_{\lambda}-\frac{k}{2R^2}\,s_{\lambda}\ ,\end{equation}
with the final equality arising from (\ref{2.8a}) and (\ref{2.8b}). Now (\ref{2.42}) is a consequence of (\ref{2.40}) and (\ref{2.43}). Since the passage from $s_a$ (and thus 
from $\sigma _a$) to $F_{ab}$ involves only first derivatives of $s_a$ we obtain for $F_{ab}$, and thus for $M_{ab}$, along $r=0$ for $k=\pm 1$ the same 
expressions as those given by (\ref{2.29a})--(\ref{2.30}) and (\ref{2.33}) in the case $k=0$. Hence summing the electromagnetic energy tensors as before with 
respect to the directions of the 3--vectors ${\bf b}$ and ${\bf n}$ yields the isotropic energy--momentum--stress tensor (\ref{2.36}) along $r=0$. This construction
can be carried out along any integral curve of the vector field $u^a$ and thus the perfect fluid (\ref{2.36}) with (\ref{2.37}) models the background electromagnetic 
radiation in all cases $k=0, \pm 1$.

\setcounter{equation}{0}
\section{Cosmic Background Gravitational Radiation}\indent
We consider now perturbations of the FLRW space--times which describe gravitational waves \cite{HE} using the gauge--invariant and covariant perturbation 
theory of Ellis and Bruni \cite{E+B}. The perturbations we are interested in here describe 
pure gravitational waves \cite{HOS} and are characterized by 
the existence of perturbations in the shear $\sigma ^{ab}$ of the matter world lines and in the matter distribution itself in the form of anisotropic stress $\pi ^{ab}$. 
The latter is a first order perturbation addition to the perfect fluid energy--momentum--stress tensor (\ref{2.6}) satisfying $\pi ^{ab}=\pi ^{ba},\ \pi ^{ab}u_b=0=\pi ^a{}_a$. 
These perturbations generate perturbations in the ``electric" and ``magnetic" parts of the Weyl conformal curvature tensor $C_{abcd}$ given respectively by \cite{E1}
\begin{equation}\label{3.1}
E_{ab}= C_{apbq}u^p\,u^q\qquad{\rm and}\qquad H_{ab}={}^*C_{apbq}u^p\,u^q\ ,\end{equation}
where ${}^*C_{apbq}=\frac{1}{2}\eta _{ap}{}^{rs}C_{rsbq}$. We note that this left dual of the Weyl tensor is equal to its right dual $C^*_{apbq}
=\frac{1}{2}\eta _{bq}{}^{rs}C_{aprs}$. The Ricci identities provide us with \cite{HOS}
\begin{equation}\label{3.2}
\sigma ^{ab}{}_{;b}=0\ ,\end{equation}together with the relationships between $E_{ab}, H_{ab}$ and $\sigma ^{ab}, \pi ^{ab}$:
\begin{equation}\label{3.3}
E_{ab}=-\dot\sigma _{ab}-\frac{2}{3}\theta\,\sigma _{ab}+\frac{1}{2}\pi _{ab}\ ,\end{equation}and
\begin{equation}\label{3.4}
H_{ab}=-\sigma_{(a}{}^{g;c}\eta _{b)fgc}u^f\ .\end{equation}The following equations are obtained from the Bianchi identities \cite{HOS}:
\begin{equation}\label{3.5}
E^{ab}{}_{;b}=-\frac{1}{2}\pi ^{ab}{}_{;b}\ ,\qquad H^{ab}{}_{;b}=0\ ,\end{equation}
\begin{eqnarray}
\dot E^{ab}+\theta\,E^{ab}+u_rH^{(a}{}_{s;d}\eta ^{b)rsd}&=&-\frac{1}{2}(\mu +p)\sigma ^{ab}-\frac{1}{2}\dot\pi ^{ab}-\frac{1}{6}\theta\pi ^{ab}\ ,\label{3.6a}\\
\dot H^{ab}+\theta\,H^{ab}-u_rE^{(a}{}_{s;d}\eta ^{b)rsd}&=&-\frac{1}{2}\eta ^{(a}{}_{rsd}\pi ^{b)sd}\,u^r\ .\label{3.6b}\end{eqnarray}

Some simple sinusoidal, monochromatic gravitational waves on a $k=0$ FLRW space--time are given by (see \cite{HOS} for families of pure 
gravitational radiation fields on isotropic cosmologies)
\begin{equation}\label{3.7}
\sigma _{ab}dx^adx^b=\lambda\,R(t)\,\delta_{\alpha\mu}\,\delta_{\beta\nu}\,\{(B^{\mu}B^{\nu}-C^{\mu}C^{\nu})\cos\phi +(B^{\mu}C^{\nu}+C^{\mu}B^{\nu})\sin\phi\}dx^{\alpha}dx^{\beta}\ ,
\end{equation}and
\begin{equation}\label{3.8}
\pi _{ab}=2\frac{\dot R}{R}\,\sigma _{ab}\ ,
\end{equation}with $\lambda, B^{\alpha}, C^{\alpha}$ and $\phi$ as in (\ref{2.20}). To verify that (\ref{3.7}) and (\ref{3.8}) satisfy (\ref{3.2})--(\ref{3.6b}) we note that 
each of the two terms on the right hand sides of (\ref{3.7}) and (\ref{3.8}) has the general form
\begin{equation}\label{3.9}
\sigma _{ab}=s_{ab}\,F(\phi )\qquad{\rm and}\qquad \pi _{ab}=\Pi _{ab}\,F(\phi )\ ,\end{equation}with $F$ arbitrary and $s_{ab}$ and $\Pi _{ab}$ symmetric, trace--free 
and orthogonal to $u^a$ on both indices. Following from (\ref{3.2})--(\ref{3.7}) the equations that $s_{ab}, \Pi _{ab}$ must satisfy are \cite{HOS}
\begin{equation}\label{3.10}
s^{ab}{}_{;b}=0\ ,\qquad \Pi ^{ab}{}_{;b}=0\ ,\end{equation}
\begin{equation}\label{3.11}
s^{ab}{}_{;c}\phi ^{,c}+\left (\frac{1}{2}\phi ^{,d}{}_{;d}-\frac{1}{3}\theta\dot\phi\right )s^{ab}=-\frac{1}{2}\dot\phi\Pi ^{ab}\ ,\end{equation}and 
\begin{equation}\label{3.12}
s^{ab;d}{}_{;d}-\frac{2}{3}\theta\dot s^{ab}+\left (p-\frac{1}{3}\mu-\frac{1}{3}\dot\theta -\frac{4}{9}\theta ^2\right )s^{ab}=-\dot\Pi ^{ab}-\frac{2}{3}\theta\Pi ^{ab}\ .\end{equation}
For future reference it is helpful to note that $E_{ab}$ and $H_{ab}$ in (\ref{3.4}) substituted into (\ref{3.6a}) results in a wave equation identical to (\ref{3.12}) but 
with $s^{ab}$ and $\Pi ^{ab}$ replaced by $\sigma ^{ab}$ and $\pi ^{ab}$ respectively. Then substitution of (\ref{3.9}) into this wave equation, and utilizing the arbitrariness 
of the function $F$, yields in particular (\ref{3.11}) and (\ref{3.12}). Now writing 
\begin{equation}\label{3.13}
s_{ab}dx^adx^b=R(t)a_{\alpha\beta}dx^{\alpha}dx^{\beta}\qquad{\rm and}\qquad \Pi_{ab}dx^adx^b=2\dot Ra_{\alpha\beta}dx^{\alpha}dx^{\beta}\ ,\end{equation}
with $a_{\alpha\beta}=a_{\beta\alpha}$ constants satisfying $a_{\alpha\beta}n^{\beta}=0=a_{\alpha\beta}\,\delta^{\alpha\beta}$ with $n^{\alpha}$ introduced in (\ref{2.17}) it is straightforward 
to verify that (\ref{3.10})--(\ref{3.12}) are satisfied.

With $\sigma _{ab}$ and $\pi _{ab}$ given by (\ref{3.7}) and (\ref{3.8}) we find using (\ref{3.3}) and (\ref{3.4}) that the only non--vanishing components of $E_{ab}$ and $H_{ab}$ 
are
\begin{eqnarray}
E_{\alpha\beta}=\lambda\,\delta_{\alpha\mu}\delta_{\beta\nu}\{(C^{\mu}C^{\nu}-B^{\mu}B^{\nu})\sin\phi +(B^{\mu}C^{\nu}+C^{\mu}B^{\nu})\cos\phi\} ,\label{3.14a}\\
H_{\alpha\beta}=\lambda\,\delta_{\alpha\mu}\delta_{\beta\nu}\{-(B^{\mu}C^{\nu}+C^{\mu}B^{\nu})\sin\phi +(C^{\mu}C^{\nu}-B^{\mu}B^{\nu})\cos\phi\}\ .\label{3.14b}\end{eqnarray}
Alternatively on account of (\ref{3.1}) the components of the perturbed Weyl tensor are now given by
\begin{equation}\label{3.15}
C_{\alpha\beta\gamma\delta}=-R^2\epsilon _{\alpha\beta\lambda}\epsilon _{\gamma\delta\sigma}E_{\lambda\sigma}\ ,\ C_{\alpha\beta\gamma 4}=R\,\epsilon _{\alpha
\beta\sigma}H_{\sigma\gamma}\ ,\ C_{\alpha 4\beta 4}=E_{\alpha\beta}\ .\end{equation}The analogue of the electromagnetic energy tensor of the previous section
 is the Bel--Robinson tensor \cite{PR} 
\begin{equation}\label{3.16}
M_{abcd}=\frac{1}{4}(C_a{}^p{}_b{}^q\,C_{cpdq}+{}^*C_a{}^p{}_b{}^q\,{}^*C_{cpdq})\ .\end{equation}
This satisfies $M_{(abcd)}=M_{abcd}$ and $M^a{}_{acd}=0$. Using (\ref{3.14a})--(\ref{3.15}) along with (\ref{2.22}) and (\ref{2.23}) we find that $M_{abcd}$ can 
be simplified to read
\begin{equation}\label{3.17}
M_{abcd}=2\lambda ^2({\bf B}\cdot{\bf B})^2\phi _{,a}\phi _{,b}\phi _{,c}\phi _{,d}\ ,\end{equation}with ${\bf B}\cdot{\bf B}$ given by (\ref{2.22}). Summing (i.e. integrating) 
this over all possible directions of ${\bf b}$ results in 
\begin{equation}\label{3.18}
M_{abcd}=\frac{64\pi\lambda ^2}{15}\phi _{,a}\phi _{,b}\phi _{,c}\phi _{,d}\ .\end{equation}Finally summing over the possible directions of ${\bf n}$ we 
arrive at an isotropic tensor ${\cal M}_{abcd}$ associated with the gravitational radiation background given by
\begin{equation}\label{3.19}
{\cal M}_{abcd}=\frac{c_1^2}{R^4}\{u_a u_b u_c u_d+2h_{(ab}u_c u_{d)}+\frac{1}{5}h_{(ab}h_{cd)}\}\ ,\end{equation}
with $c_1^2=256\pi ^2\lambda ^2/15$.

To extend this result to the FLRW space--times with $k=\pm 1$ we follow the pattern of the electromagnetic case described in the previous section. It will be 
sufficient to establish (\ref{3.19}) along the world line $r=0$. By analogy with the electromagnetic case we might expect that the solution (\ref{3.7}) and (\ref{3.8}) might hold 
along $r=0$ in the case $k=\pm 1$ but this is not the case. The anisotropic stress perturbation (\ref{3.8}) can be retained but the matter shear perturbation has to be 
adjusted by the inclusion of a factor $f^{-7/3}$ where $f$ is given by (\ref{2.1a}). The simplest way to see this is in terms of the variables (\ref{3.13}). With or without the 
factor $f^{-7/3}$ multiplying $s_{ab}$ the field equations will be satisfied along $r=0$ except the final wave equation. This is because although $s^{ab}{}_{;b}=0$ along 
$r=0$ and the derivative of the left hand side here along $r=0$ (i.e. in the direction $u^a$) also vanishes the spatial derivative $h^d_f(s^{ab}{}_{;b})_{;d}$ does not vanish 
along $r=0$, analogous to (\ref{2.40}). The wave equation evaluated along $r=0$ now instead of (\ref{3.12}) reads
\begin{equation}\label{3.20}
s^{ab;d}{}_{;d}-\frac{2}{3}\theta\dot s^{ab}+\left (p-\frac{1}{3}\mu-\frac{1}{3}\dot\theta -\frac{4}{9}\theta ^2\right )s^{ab}-\frac{3}{2}(s^{d(a}{}_{;d})^{;b)}=-\dot\Pi ^{ab}-\frac{2}{3}\theta\Pi ^{ab}\ .\end{equation}
If we put $s_{\alpha\beta}=f^n\,R(t)\,a_{\alpha\beta}$ (with a view to determining the real exponent $n$) and $\Pi _{\alpha\beta}=2\dot R\,a_{\alpha\beta}$ then along $r=0$ we find that 
\begin{equation}\label{3.21}
(s^d_{\alpha ;d})_{;\beta}=\frac{(n-1)k}{2R^2}s_{\alpha\beta}\ ,\end{equation} and (\ref{3.20}) reads explicitly
\begin{equation}\label{3.22}
s_{\alpha\beta ;d}{}^{;d}-\left (\frac{\ddot R}{R}+\frac{3\dot R^2}{R^2}+\frac{2k}{R^2}\right )s_{\alpha\beta}-\frac{3k}{4R^2}(n-1)s_{\alpha\beta}=0\ ,\end{equation}
along $r=0$. Direct calculation of the first term here, using the formulas in the appendix, yields
\begin{equation}\label{3.23}
s_{\alpha\beta ;d}{}^{;d}=\frac{3nk}{2R^2}s_{\alpha\beta}+\left (\frac{\ddot R}{R}+\frac{3\dot R^2}{R^2}+\frac{3k}{R^2}\right )s_{\alpha\beta}\ ,\end{equation}
along $r=0$. Now (\ref{3.23}) in (\ref{3.22}) results in the requirement that $n=-7/3$. Since the perturbed Weyl tensor described by $E_{ab}, H_{ab}$ given by (\ref{3.3}) and 
(\ref{3.4}) only involves first derivatives of $\sigma _{ab}, \pi _{ab}$ the perturbed Weyl tensor calculated along $r=0$ will in the case $k=\pm 1$ be given again by (\ref{3.14a}) 
and (\ref{3.14b}), leading once again to (\ref{3.19}) along $r=0$ and thus, by isotropy, along any integral curve of the vector field $u^a$.

\setcounter{equation}{0}
\section{Energy--Momentum--Stress Tensor for Gravitational Background}\indent
The energy--momentum--stress tensor ${\cal M}_{ab}$ in (\ref{2.36}) describes macroscopically the CMB radiation. It has been derived starting from an 
electromagnetic energy tensor $M_{ab}$ in (\ref{2.31}). ${\cal M}_{ab}$ has the same algebraic symmetries as $M_{ab}$ in (\ref{2.31}) (symmetric and trace--free) but ${\cal M}_{ab}$ is \emph{not} 
an electromagnetic energy tensor. If it were, and the corresponding electric and magnetic fields $E_a$ and $H_a$ were given in general by (\ref{2.11}), then they would 
have to satisfy the algebraic equations
\begin{equation}\label{4.1}
E_aH_b=E_bH_a\ ,\end{equation}
and
\begin{equation}\label{4.2}
E_a\,E_b+H_a\,H_b=\frac{1}{3}(E_c\,E^c+H_c\,H^c)\,h_{ab}\ ,\end{equation}
because ${\cal M}_{ab}$ includes no energy flux relative to $u^a$ and no anisotropic stress (see \cite{E1}, p.14). These equations imply that
$E_a=0=H_a$. In similar fashion the tensor ${\cal M}_{abcd}$ in (\ref{3.19}) has the same symmetries as the Bel--Robinson tensor (\ref{3.16}) (symmetric and trace--free on 
any pair of indices) but it is not a Bel--Robinson tensor. If it were a Bel--Robinson tensor  then it would be derivable from electric and magnetic parts $E_{ab}$ 
and $H_{ab}$ of a Weyl tensor given in general by (\ref{3.1}). These parts would then have to satisfy analogous equations to (\ref{4.1}) and (\ref{4.2}) namely,
\begin{equation}\label{4.3}
H_{ab}\,E_{cd}+H_{cb}\,E_{ad}=H_{ad}\,E_{cb}+H_{cd}\,E_{ab}\ ,\end{equation}
and
\begin{equation}\label{4.4}
E_{ab}\,E_{cd}+H_{ab}\,H_{cd}=\frac{Q}{30}\,(3\,h_{ac}\,h_{bd}+3\,h_{ad}\,h_{bc}-2\,h_{ab}\,h_{cd})\ ,\end{equation}
with $Q=E_{ab}\,E^{ab}+H_{ab}\,H^{ab}$ (see \cite{AG}, eqns.(5.1) and (5.2)). But these equations imply that $E_{ab}=0=H_{ab}$.

In this paper we construct isotropic (with respect to the congruence tangent to $u^a$) energy--momentum--stress tensors by summing over waves propagating in all directions. In this context it is perhaps helpful to 
draw attention to the fact that no similarly isotropic electromagnetic field, described by $E_a$ and $H_a$, exists (since these vectors are orthogonal to $u^a$) but tensors quadratic in these quantities such as $E_aE_b$ have non--vanishing isotropic forms which are proportional to the projection tensor defined following (\ref{2.3}). A similar comment applies to the tensor fields $E_{ab}$ and $H_{ab}$.

${\cal M}_{abcd}$ in (\ref{3.19}) is a gauge--invariant tensor field on the FLRW space--times which is small of  second order (since it is quadratic in first order perturbations of the FLRW space--times), has 
dimensions of $({\rm length})^{-4}$ and has divergence given by
\begin{equation}\label{4.5}
{\cal M}^{abcd}{}_{;d}=\frac{2\,c_1^2}{3\,R^4}\,\theta\,\left\{u^au^bu^c+\frac{1}{3}(h^{ab}u^c+h^{bc}u^a+h^{ac}u^b)\right\}\ .\end{equation}On the other hand we find that
\begin{equation}\label{4.6}
(u_au_b{\cal M}^{abcd})_{;d}=0\ ,\end{equation}where by (\ref{3.19})
\begin{equation}\label{4.7}
u_au_b{\cal M}^{abcd}=\frac{c_1^2}{R^4}(u_cu_d+\frac{1}{3}h_{cd})\ .\end{equation}This tensor is symmetric, trace--free and divergence--free. It has the attributes of an energy--momentum--stress 
tensor except for the fact that its dimensions are $({\rm length})^{-4}$. However using the scale factor $R(t)$ we can define an energy--momentum--stress tensor with dimensions 
$({\rm length})^{-2}$ by multiplying (\ref{4.7}) by $R_0^2$ to give
\begin{equation}\label{4.8}
{\cal T}^{ab}=\mu _g\,u^a\,u^b+p_g\,h^{ab}\ ,\end{equation}
with
\begin{equation}\label{4.9}
p_g=\frac{1}{3}\mu _g=\frac{c_1^2R_0^2}{3R^4}\ ,\end{equation}for some $R_0=R(t_0)\neq 0$ and the subscripts on $p$ and $\mu$ reflect the gravitational wave origin of these quantities. In this 
way \emph{we associate an energy--momentum--stress tensor of the CMB type with an isotropic background of gravitational radiation}. It is a second order perturbation of a background isotropic 
energy--momentum--stress tensor such as that describing the CMB, neglecting anisotropies. It would be interesting to incorporate it into the Ellis--Bruni perturbation theory at second order. The 
energy--momentum--stress tensor (\ref{4.8}) will act as a source of second order perturbations of the isotropic cosmologies. Anisotropies in this cosmic 
background \emph{gravitational} radiation will thus be third order.

\appendix
\section{Useful Riemannian Connection Components} \setcounter{equation}{0}
The non--vanishing components of the Riemannian connection associated with the metric tensor $g_{ab}$ given via the line--element (\ref{2.1}) with 
(\ref{2.1a}) are
\begin{equation}\label{A1}
-\Gamma ^1_{11}=\Gamma ^1_{22}=\Gamma ^1_{33}=f^{-1}\frac{\partial f}{\partial x}\ , \Gamma ^1_{12}=-f^{-1}\frac{\partial f}{\partial y}\ ,
\Gamma ^1_{13}=-f^{-1}\frac{\partial f}{\partial z}\ ,\end{equation}
\begin{equation}\label{A2}
-\Gamma ^2_{22}=\Gamma ^2_{11}=\Gamma ^1_{33}=f^{-1}\frac{\partial f}{\partial y}\ , \Gamma ^2_{12}=-f^{-1}\frac{\partial f}{\partial x}\ ,
\Gamma ^2_{23}=-f^{-1}\frac{\partial f}{\partial z}\ ,\end{equation}
\begin{equation}\label{A3}
-\Gamma ^3_{33}=\Gamma ^3_{11}=\Gamma ^3_{22}=f^{-1}\frac{\partial f}{\partial z}\ , \Gamma ^3_{13}=-p_o^{-1}\frac{\partial f}{\partial x}\ ,
\Gamma ^3_{23}=-f^{-1}\frac{\partial f}{\partial y}\ ,\end{equation}
\begin{equation}\label{A4}
\Gamma ^4_{\alpha\beta}=R\,\dot R\,f^{-2}\,\delta _{\alpha\beta}\ ,\end{equation}
and
\begin{equation}\label{A5}
\Gamma ^{\alpha}_{\beta 4}=\frac{\dot R}{R}\,\delta _{\alpha\beta}\ .\end{equation}From these we see that along $r=0$ the only non--vanishing components 
are $\Gamma ^4_{\alpha\beta}=R\,\dot R\,\delta _{\alpha\beta}$ and $\Gamma ^{\alpha}_{\beta 4}=\dot R/R\,\delta _{\alpha\beta}$. A further useful consequence is that 
along $r=0$ 
\begin{equation}\label{A6}
\Gamma ^{\alpha}_{\beta\sigma ,\sigma}=-\frac{3k}{2}\,\delta _{\alpha\beta}\ .\end{equation}

\end{document}